\documentclass[twocolumn,floats,showpacs,superscriptaddress,pre]{revtex4}
\usepackage{amsmath,amssymb,bm}
\usepackage{graphics}
\usepackage{epsfig}
\usepackage{graphicx}

\input{tcilatex}

\begin{document}

\title{On a New Form of Quantum Mechanics}
\author{N. N. Gorobey and A. S. Lukyanenko}
\email{alex.lukyan@rambler.ru}
\affiliation{Department of Experimental Physics, St. Petersburg State Polytechnical
University, Polytekhnicheskaya 29, 195251, St. Petersburg, Russia}

\begin{abstract}
We propose a new form of nonrelativistic quantum mechanics which is based on
a quantum version of the action principle.
\end{abstract}

\maketitle
\date{\today }





\section{\textbf{INTRODUCTION}}

Quantum mechanics was originated in two principally different forms: the
matrix mechanics of Heisenberg and the wave mechanics of Schr\"{o}dinger.
Both forms are equivalent in general and are the complement of one another.
Latter Feynman \cite{F} proposed a third formulation in terms of a path
integral which is equivalent to the previous two. The use of any alternative
formulation is obvious: it opens new possibilities in the development of the
theory \cite{F1}. Following the logic of the quantum theory development one
can suppose that not all alternative formulations are found. In this paper
we propose a new formulation of the non-relativistic quantum mechanics with
a quantum version of the action principle in a ground. We shall come to this
formulation pushing from the Schr\"{o}dinger wave theory. Then the canonical
foundation of the quantum action principle will be done.

\section{\textbf{WAVE FUNCTIONAL}}

For simplicity we shall consider here the dynamics of a particle of mass $m$
in the one-dimensional space. In the Schr\"{o}dinger theory a quantum state
of a particle is described by a wave function $\psi \left( x,t\right) $.
This wave function has the meaning of the probability density to observe the
particle nearby a point with a coordinate $x$ at a moment of time $t$ if the
following normalization condition is fulfilled: \ \ \ \ \ \ \ \ \ \ \ \ \ \
\ \ \ \ \ \ \ 
\begin{equation}
\overset{+\infty }{\underset{-\infty }{\dint }}\left| \psi \left( x,t\right)
\right| ^{2}dx=1.  \label{1}
\end{equation}
\ \ \ \ \ \ \ \ \ \ \ \ \ \ \ \ \ \ \ \ \ \ \ \ The evolution of a quantum
state is described by the Schr\"{o}dinger equation: \ \ \ \ \ \ \ \ \ \ \ \
\ \ 
\begin{equation}
i\hbar \overset{\cdot }{\psi }=-\frac{\hbar ^{2}}{2m}\psi ^{\prime \prime
}+U\left( x,t\right) \psi ,  \label{2}
\end{equation}
where the dot denotes the partial derivative on time and the hatch denotes
the partial derivative on $x$-coordinate. The normalization condition (\ref%
{1}) is conserved during the Schr\"{o}dinger evolution. The Schr\"{o}dinger
equation (\ref{2}) may be obtained as the Euler-Lagrange equation for the
action:

\begin{eqnarray}
I_{S}\left[ \psi \right] &\equiv &\overset{T}{\underset{0}{\dint }}dt\overset%
{+\infty }{\underset{-\infty }{\dint }}dx\left[ \frac{1}{2}i\hbar \left( 
\overline{\psi }\overset{\cdot }{\psi }-\overset{\cdot }{\overline{\psi }}%
\psi \right) \right.  \notag \\
&&\left. -\frac{\hbar ^{2}}{2m}\overline{\psi }^{\prime }\psi ^{\prime }-U%
\overline{\psi }\psi \right] .  \label{3}
\end{eqnarray}

The wave function describes a quantum state of the particle at each moment
of time. Let us introduce a new object -- a function of a trajectory of the
particle, i.e., a functional $\Psi \left[ x\left( t\right) \right] $ which
describes the particle dynamics on a whole time interval $\left[ 0,T\right] $%
. It will be called \emph{the wave functional}. For this purpose, let us
divide the interval $\left[ 0,T\right] $\ by points $t_{n}$\ on $N$\ small
parts of equal length $\varepsilon =T$ $/N$ . Let us approximate a
trajectory $x=x\left( t\right) $\ of the particle by a broken line with
vertices $x_{n}=x\left( t_{n}\right) $. We define the wave functional as a
product: 
\begin{equation}
\Psi \left[ x\left( t\right) \right] \equiv \underset{n}{\dprod }\psi \left(
x_{n},t_{n}\right) .  \label{4}
\end{equation}
The normalization condition (\ref{1}) may be rewritten as:

\begin{equation}
\left( \Psi ,\Psi \right) \equiv \dint \underset{n}{\dprod }dx_{n}\left\vert
\Psi \left[ x\left( t\right) \right] \right\vert ^{2}=1.  \label{5}
\end{equation}
In fact, the wave functional is a function of $N+1$\ variables:

\begin{equation}
\Psi \left[ x\left( t\right) \right] \equiv F\left(
x_{0},...,x_{n},...,x_{N}\right) ,  \label{6}
\end{equation}
but the limit $N\rightarrow \infty $\ is assumed. Taking into account this
limit, let us define the variation derivative of the functional (6) as
follows:

\begin{equation}
\frac{\delta \Psi \left[ x\left( t\right) \right] }{\delta x\left(
t_{n}\right) }\equiv \frac{1}{\varepsilon }\frac{\partial F}{\partial x_{n}}=%
\frac{1}{\varepsilon }\frac{\partial \psi \left( x_{n},t_{n}\right) }{%
\partial x_{n}}\underset{n^{\prime }\neq n}{\dprod }\psi \left( x_{n^{\prime
}},t_{n^{\prime }}\right) .  \label{7}
\end{equation}

Let us consider the integral defined by the corresponding integral sum:

\begin{equation}
\underset{0}{\overset{T}{\dint }}dt\overset{\cdot }{x}\left( t\right) \frac{%
\delta \Psi \left[ x\left( t\right) \right] }{\delta x\left( t\right) }\cong 
\underset{n}{\dsum }\Delta x_{n}\frac{1}{\varepsilon }\frac{\partial F}{%
\partial x_{n}},  \label{8}
\end{equation}
where $\Delta x_{n}\equiv x_{n+1}-x_{n}$. The integral (\ref{8}) has the
meaning of a variation of the wave functional (\ref{4}) which is originated
by an infinitesimal shift of the broken line ``back''\ in time. This shift
is represented as the successive replacement of vertices: $x_{n}\rightarrow
x_{n+1}$. On the other hand, this variation may be considered as the result
of the time shift on the one step back: $t\rightarrow t-\varepsilon $. Then
one obtains:

\begin{equation}
\underset{0}{\overset{T}{\dint }}dt\overset{\cdot }{x}\left( t\right) \frac{%
\delta \Psi \left[ x\left( t\right) \right] }{\delta x\left( t\right) }\cong 
\underset{n}{-\dsum }\frac{\partial \psi \left( x_{n},t_{n}\right) }{%
\partial t}\underset{n^{\prime }\neq n}{\dprod }\psi \left( x_{n^{\prime
}},t_{n^{\prime }}\right) .  \label{9}
\end{equation}

\section{\textbf{QUANTUM ACTION PRINCIPLE}}

The action (\ref{3}), which defines the Schr\"{o}dinger dynamics in terms of
the wave function $\psi \left( x,t\right) $, can be transformed in a
functional of the wave functional $\Psi \left[ x\left( t\right) \right] $.
The following formula,

\begin{eqnarray}
I_{S}\left[ \psi \right] &=&-\underset{0}{\overset{T}{\dint }}dt\dint 
\underset{t}{\dprod }dx\left\{ \frac{1}{2}i\widetilde{\hbar }\overset{\cdot }%
{x}\left( t\right) \left[ \overline{\Psi }\frac{\delta \Psi }{\delta x\left(
t\right) }-\frac{\delta \overline{\Psi }}{\delta x\left( t\right) }\Psi %
\right] \right.  \notag \\
&&\left. +\frac{\hbar ^{2}}{2m}\frac{\delta \overline{\Psi }}{\delta x\left(
t\right) }\frac{\delta \Psi }{\delta x\left( t\right) }+U\left( x\left(
t\right) ,t\right) \overline{\Psi }\Psi \right\} ,  \label{10}
\end{eqnarray}
is a ``bridge''\ between the Schr\"{o}dinger and our representations. For
the multiplicative functional (\ref{4}), which is related with the $%
\varepsilon $-division of the interval of time $\left[ 0,T\right] $, the
constant $\widetilde{\hbar }$\ is

\begin{equation}
\widetilde{\hbar }\equiv \varepsilon \hbar .  \label{11}
\end{equation}
The proof of the formula (\ref{10}) is based on the definition (\ref{7}) of
the variation derivative of the wave functional, the equation (\ref{9}) and
the normalization condition for the wave function (\ref{1}) which is
conserved in time.

The new representation of the Schr\"{o}dinger action in terms of the wave
functional gives us a possibility for an alternative formulation of quantum
dynamics. The new formulation is based on the following equation:

\begin{eqnarray}
\widehat{I}\Psi &\equiv &\overset{T}{\underset{0}{\dint }}dt\left[ \frac{%
\widetilde{\hbar }}{i}\overset{\cdot }{x}\left( t\right) \frac{\delta \Psi }{%
\delta x\left( t\right) }+\frac{\widetilde{\hbar }^{2}}{2m}\frac{\delta
^{2}\Psi }{\delta x^{2}\left( t\right) }-U\left( x\left( t\right) ,t\right)
\Psi \right]  \notag \\
&=&\lambda \Psi .  \label{12}
\end{eqnarray}
Here $\lambda $\ is an eigenvalue of the operator $\widehat{I}$ which is a
quantum version of the classical canonical action:

\begin{equation}
I\left[ x\left( t\right) ,p\left( t\right) \right] \equiv \overset{T}{%
\underset{0}{\dint }}dt\left[ p\overset{\cdot }{x}-\frac{p^{2}}{2m}-U\left(
x\left( t\right) ,t\right) \right] .  \label{13}
\end{equation}
The \textquotedblleft quantization\textquotedblright\ of the classical
action (\ref{13}) is performed by the replacement of the canonical momentum $%
\ p\left( t\right) $\ by the functional-differential operator:

\begin{equation}
\widehat{p}\left( t\right) \equiv \frac{\widetilde{\hbar }}{i}\frac{\delta }{%
\delta x\left( t\right) }.  \label{14}
\end{equation}

Let us formulate the quantum action principle as a search for an extremum in
the set of eigenvalues $\lambda $, assuming that $\lambda $\ depends on a
certain set of continuous parameters. According to (\ref{10}), on the set of
the multiplicative functionals (\ref{4}) we have:

\begin{equation}
\lambda =\left( \Psi ,\widehat{I}\Psi \right) =I_{S}\left[ \psi \right] ,
\label{15}
\end{equation}
Therefore, the quantum action principle reduces to the well-known action
principle of the Schr\"{o}dinger wave mechanics if only multiplicative wave
functionals are considered.

\section{\textbf{NEW FORM OF CANONICAL QUANTIZATION}}

The canonical foundation of the new form of quantum dynamics consists in the
definition of new rules of transition from classical to quantum mechanics.
Old rules are a ``quantum deformation''\ of classical dynamics which is
formulated in the canonical form by use of the Poisson brackets (PB) \cite%
{FD}. These brackets are defined as the Lie brackets on the set of functions
of canonical variables which obey the canonical relation (in the case of one
dimension space):

\begin{equation}
\left\{ x,p\right\} =1,  \label{16}
\end{equation}
see, for example, in Ref. \cite{M}. The relation (\ref{16}) is defined in a
certain moment of time but the classical dynamics conserves PB-relations in
time.

A modification of the classical dynamics proposed here consists in the
replacement of the ordinary PB by a new ones which obey a non-simultaneous
canonical relation (all others are equal\ to zero):

\begin{equation}
\left\{ x\left( t\right) ,p\left( t^{\prime }\right) \right\} =\delta \left(
t-t^{\prime }\right) .  \label{17}
\end{equation}
The new definition of PB permits us to formulate classical equations of
motion as the PB-relations:

\begin{equation}
\left\{ x\left( t\right) ,I\right\} =\left\{ p\left( t\right) ,I\right\} =0,
\label{18}
\end{equation}
where $I$\ is the classical action (\ref{13}). The equations (\ref{18}) are
conditions of extremum of the classical action.

The canonical quantization of this theory consists in the replacement of
classical canonical variables by operators and the replacement of canonical
PB-relations by the corresponding commutation relations \cite{FD}. In our
case the commutator

\begin{equation}
\left[ \widehat{x}\left( t\right) ,\widehat{p}\left( t^{\prime }\right) %
\right] =i\widetilde{\hbar }\delta \left( t-t^{\prime }\right)  \label{19}
\end{equation}
defines a quantum algebra of the canonical variables, where $\widetilde{%
\hbar }$ is a new ``Plank''\ constant with the dimensionality $Dj\cdot s^{2}$%
. The canonical commutation relation (\ref{19}) is valid if we consider, as
usually, $\widehat{x}\left( t\right) $ as the operator of product by $%
x\left( t\right) $\ and $\ \widehat{p}\left( t\right) $ as the operator of
the functional differentiation (\ref{14}) in the space of wave functionals $%
\Psi \left[ x\left( t\right) \right] $. After that, the quantum version of
the action principle formulated in this work arises naturally.

\section{\textbf{CONCLUSIONS}}

The equation (\ref{12}) is the main result of our work. It is an analog of
the Schr\"{o}dinger equation in the new formulation of quantum mechanics.
The new formulation is equivalent to the Schr\"{o}dinger theory, but it
opens new possibilities for the development of quantum theory. The
description of processes of birth and annihilation of particles without the
use of the secondary quantization formalism will be one of the new
possibilities.

We are thanks V. A. Franke and A. V. Goltsev for useful discussions.




\end{document}